# Search for relationship between duration of the extended solar cycles and amplitude of sunspot cycle

**A.G. Tlatov**
*Kislovodsk Solar Station of the Pulkovo Observatoy, Kislovodsk, Russia*
*E-mail: solar@narzan.com*

**Abstract.** Duration of the extended solar cycles is taken into the consideration. The beginning of cycles is counted from the moment of polarity reversal of large-scale magnetic field in high latitudes, occurring in the sunspot cycle *n* till the minimum of the cycle *n+2*. The connection between cycle duration and its amplitude is established. Duration of the "latent" period of evolution of extended cycle between reversals and a minimum of the current sunspot cycle is entered. It is shown, that the latent period of cycles evolution is connected with the next sunspot cycle amplitude and can be used for the prognosis of a level and time of a sunspot maximum. The $24^{th}$ activity cycle prognosis is done. Long-term behavior of extended cycle's lengths is considered.

## 1. Introduction

Duration of activity cycle is an important parameter necessary for understanding of the solar cyclicity. In order to define the sunspots cycle length, as a rule, the moments of a sunspots minimum are used. Average duration of sunspots cycle duration is near ~11 years, but can vary from ~8 till 15 years. Between sunspots cycle's length and its amplitude there is a relation: $W_{max}=379,9$ (±64,5)- 24.959 (±64,5) ·L, where L and $W_{max}$ show the length and maximal Wolf number (Chistaykov, 1997). At the same time the duration of a sunspots cycle can differ from the duration of an activity cycle (Harvey, 1992) and the duration of a large-scale magnetic fields cycle (Makarov et al., 2003).

For the description of time responses of a solar magnetic cycle it is possible to use the moments of a polarity reversal of a large-scale magnetic field. In the paper (Makarov et al., 2001) it has been shown, that speed of drift of boundary polarity reversal line to poles depends on the sunspots area sum in current cycle. The time between an epoch of a minimum and the moments of a polarity reversal on poles also depends on amplitude of the current sunspots cycle (Makarov et al., 2003).

In this paper the connection between the moments of a polarity reversal of a large-scale magnetic field in polar areas and the amplitude of the next cycle of activity is investigated. Along with that long-term changes of duration of large-scale magnetic field cycles and their link with long-term variations of activity and a climate are surveyed.

## 2. Duration of the extended activity cycles

Evolution of a large-scale magnetic field in various latitudes can be investigated on series of synoptic Hα charts. The polarity boundary lines of magnetic field are represented on these charts. As markers of neutral lines the optical observations data are used. These are observations of filaments, channels of filaments or prominences. Now summary number Hα charts covers the period since 1887 till now, and is commensurable with the length of some sunspots groups (McIntosh, 1979; Makarov and Sivaraman, 1989; Vasil'eva, 1998).

The distribution of a large-scale magnetic field has a specific organization in the latitude called zone structure. During the epoch close to a sunspots maximum there is a change of a sign of a large-scale magnetic field on poles. This moment is important for the formation of the new poloidal magnetic fields of the Sun.

Let's consider a time interval between the moment of a polarity reversal according to the data of Hα maps and the following moment of a solar activity minimum $\Delta T_R = T_{min}^{n+1} - T_{H\alpha}^n$. In table 1 the moments of a polarity reversal presented on Hα charts, taken from paper (Makarov, Makarova, 1996) specified and supplemented by the latest data. In this table the moments of a polarity reversal for northern $T_N$ and southern $T_S$ hemispheres are presented. In case of single-fold polar magnetic field reversals the moment for a hemisphere where it passed later was chosen. In some cycles, three-fold reversal polar field was observed. In case of thrice-repeated moments of polarity reversals $T_{H\alpha}$ instanced in table 1 are taken. On fig. 1 The link between an interval $\Delta T_R$ with amplitude of the next sunspot cycle is presented. Function of a regression can be expressed as:

$$W_{max}^{n+1} = 320(\pm51) - 38,2(\pm9,6) \cdot \Delta T_R, r = 0,78; \sigma = 27. \quad (1)$$

Apparently, that the shorter the time interval $\Delta T_R$ is, the higher is the amplitude of the next sunspots cycle. There is also a connection (a Fig. 2) between duration from the polarity reversal moment in cycle $n$ and the moment of a sunspot maximum $T_{max}^{n+1}$ in a new cycle $\Delta T_2 = T_{max}^{n+1} - T_{H\alpha}^n$ with amplitude of a new sunspot cycle $W_{max}^{n+1}$ which can be presented as:

$$W_{max}^{n+1} = 352(\pm40) - 24,9(\pm4,3) \cdot \Delta T_2, r = 0,87; \sigma = 21. \quad (2)$$

It is possible to enter one more time interval, equal to an interval between the moment of a polarity reversal of a large-scale magnetic field $T_{H\alpha}$ in a cycle n and the moment of a sunspots minimum of a cycle n+2: $\Delta T_3 = T_{min}^{n+2} - T_{H\alpha}^n$. This period varies from 14,2 years for 18-th cycle till 18,6 years for 14[th] activity cycle (a Fig. 3). Between the amplitude of a sunspot cycle and an interval $\Delta T_3$ there is a relationship:

$$W_{max}^{n+1} = 450(\pm101) - 20,7(\pm6,4) \cdot \Delta T_3, r = 0,72; \sigma = 31. \quad (3)$$

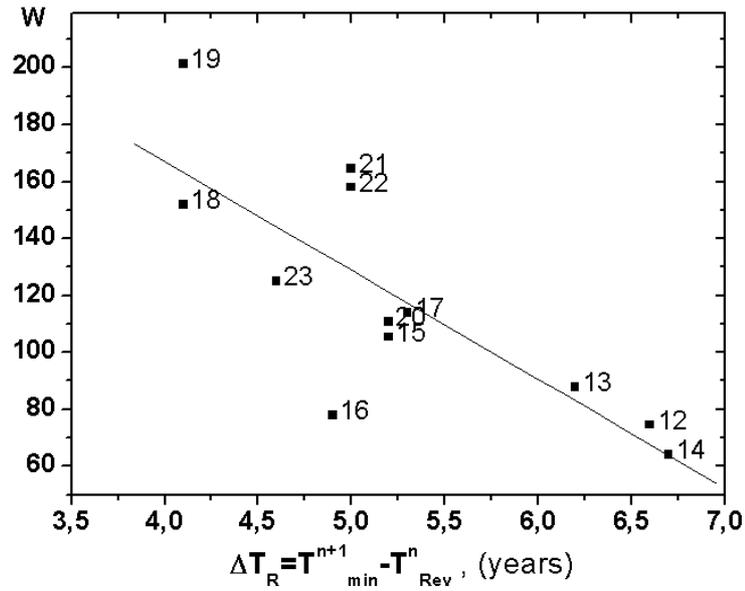

**Fig. 1.** Amplitude of an sunspots index as the function from duration of time interval $\Delta T_R$ between the polarity reversal moments of a large-scale magnetic field in a cycle *n* and a minimum of a cycle of activity *n+1*.

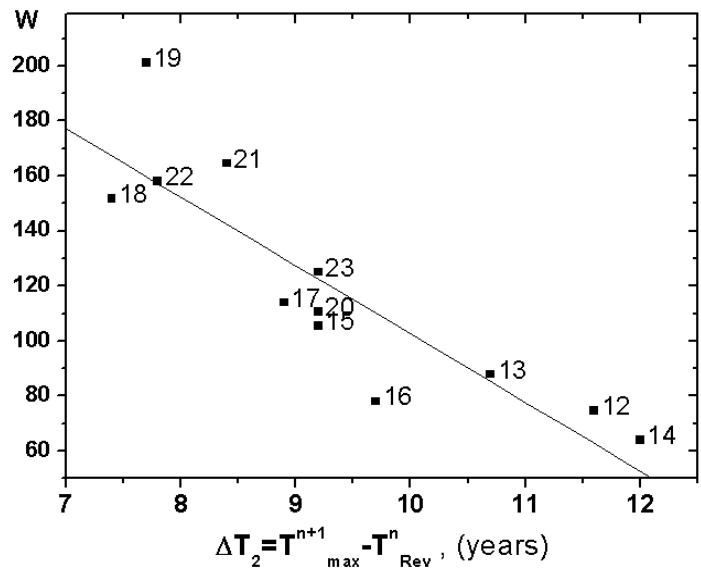

**Fig. 2.** Amplitude of the sunspots index as a function from the time interval $\Delta T_2$ between the moments of a polarity reversal in a cycle *n* and a maximum of a cycle of activity *n+1*.

Between the interval $\Delta T_2$ and duration of the interval $\Delta T_R$ there is a link:
$$\Delta T_R^{max} = 1{,}1(\pm 1{,}3) + 1{,}54(\pm 0{,}24) \cdot \Delta T_R, r = 0{,}88; \quad \sigma = 0{,}7, \quad (4)$$
here time intervals are expressed in years (Fig. 4).

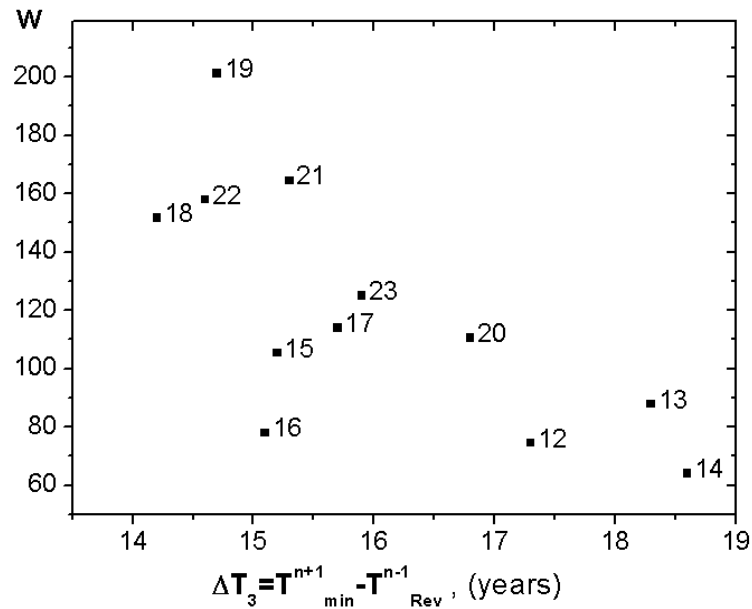

**Fig. 3.** Amplitudes of the sunspots cycles as a function from the overall duration of extended activity cycles $\Delta T_3$, counted from the moment of the polarity reversal in a cycle n-1 before a minimum of a sunspots cycle n+1.

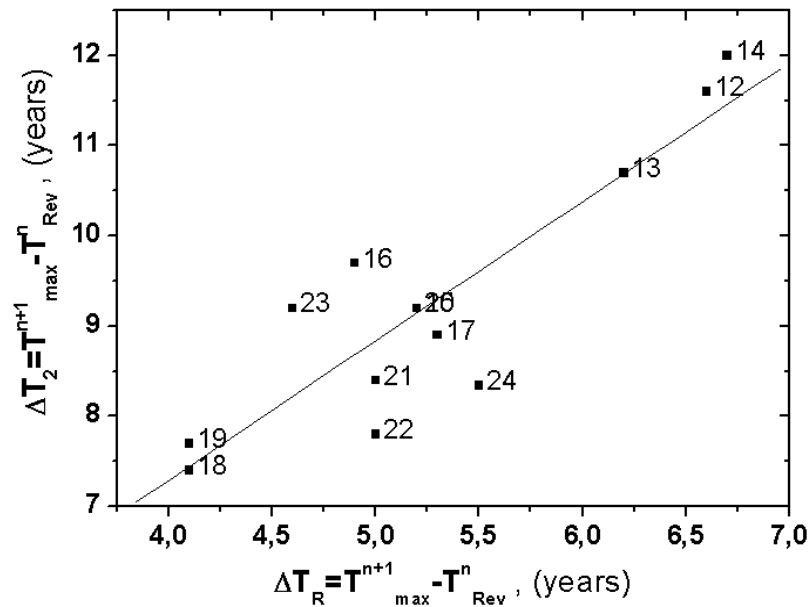

**Fig. 4.** Connection between the time interval of a between a polarity reversal and a minimum of activity $\Delta T_R$ with an interval between a polarity reversal and approach of a maximum of sunspots activity $\Delta T_2$.

## 3. Long-term variations of duration of cycles of a large-scale magnetic field

The large-scale magnetic field of the Sun as well as sunspots cycles has long-term modulation (Makarov et al., 2002). Duration of cycles of a large-scale magnetic field varied within the past century.

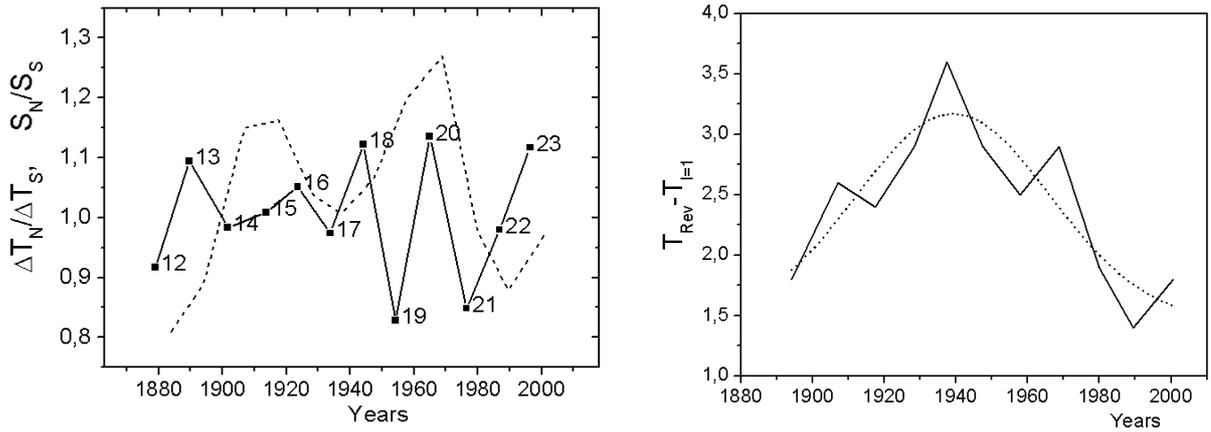

**Fig. 5.** (left) The relation of time intervals between the polarity reversals in high latitudes for the northern and southern hemispheres (solid line). The relation of the sums of the sunspots areas in cycles for the northern and southern hemispheres (dashed line).

**Fig. 6.** (right) The time interval between the moments of a polarity reversal of the dipole components of large-scale magnetic field $T_{l=1}$ and the moment of a polarity reversal in high latitudes $T_{H\alpha}$. The envelope line (dot line) is drawn.

On fig. 5 the ratio of times intervals between the polarity reversal moments in northern and southern hemispheres are presented. For comparison on the diagram the ratio of the sunspots area sums in cycles for northern and southern hemispheres also is put. With respect to the duration of polarity reversals 22 years modulation which can be connected with presence relict poloidal magnetic field is well appreciable.

For analysis of duration of a large-scale magnetic field alongside with the use of the moment of a polarity reversal other characteristics of a large-scale magnetic field can be used, for example the moments of reversal of dipole components of magnetic field $T_{l=1}$ (Tab. 1).

On fig. 6 changes of a time interval between the polarity reversal moment on high latitudes and the moment of a polarity reversal dipole component of a large-scale magnetic field are presented. $\Delta T = T_{H\alpha} - T_{l=1}$. There is a long-term trend of this time interval with a maximum near to ~1940 year.

## 4. Conclusion

The discovered links between the time intervals counted off the moment of a large-scale magnetic field reversal and the amplitude of the next sunspot cycle show that the solar activity cycle has the longer duration, than an 11-years sunspot cycle. Expressions (1) - (3) show the presence of the connection between duration of a "extended" cycle (Wilson et al., 1988) beginning from the polarity reversal moment in polar areas of the Sun and the amplitude of activity of the following sunspots cycle.

Thus the time interval $\Delta T_3$ can be interpreted as the duration of the extended activity cycle. The time interval $\Delta T_R$ between the polarity reversal and a minimum of activity in *n+1* can be regarded as the "latent" period of evolution of a extended solar cycle. This time interval can be used for the prognosis of the amplitude of the next sunspot cycle.

If to accept the minimum of 24[th] cycle of activity has come in 2007,2 the amplitude of 24[th] cycle can be estimated as ~110 (±27) W. The ratio (4) enables estimations of approach of a maximum of a new cycle of activity. Maximum of 24[th] activity cycle is possibly expected in 2011,3 (±0,7) year.

It should be pointed out that the correlation (3) shows that during the period of deep minima of activity, for example during Maunder minimum, the duration of an interval $\Delta T_3$ is close to 22 years, that is duration of the extended activity cycle during this period tends to the duration of a solar magnetic cycle.

Determination of the role of the moment of a polarity reversal in polar areas and duration of the activity cycle of about 15-18 years corresponds to a transport dynamo model of generation of solar cyclicity with the closed meridional circulation (Tlatov, 1995, 1997; Choudhuri et al., 1995). Probably, changes in duration of the extended activity cycle in the interval of 14-18 years are linked with the speed of drift of a generation wave at the bottom of the convective zone is a variable quantity which can be calculated from parameter $\Delta T_3$, varies in limits v~1,3-1,7 m/s.

Long-term variations of cycles of a large-scale magnetic field are coupled with 22-year-old and secular variations of the solar activity

This paper was supported by the Russian Fund of Basic Researches, projects 06-02-16333

**Table 1.** Duration of time intervals between a polarity reversal of large-scale magnetic field $T_{H\alpha}$, received from the moments of polarity reversals in the northern and southern hemisphere $T_N$ and $T_S$. In case of triple polarity reversals the moments of the first and the third wave are given. Also the moments of a polarity reversal dipole builders of large-scale magnetic field $T_{l=1}$ in relation to the moments of minimum $T_{min}$ of activity of sunspots are presented.

| No | $T_{min}$ | $T_N$ | $T_S$ | $T_{H\alpha}$ | $T_{min} - T_{H\alpha}$ | $T^{n+1}_{min} - T_{H\alpha}$ | $T_{l=1}$ |
|---|---|---|---|---|---|---|---|
| 11 | 1867.2 | | | | | | |
| 12 | 1878.9 | 1872,3 | 1872,3 | 1872,3 | 6,6 | 17,3 | |
| 13 | 1889.6 | 1883,4 | 1883.7<br>1885,8 | 1883,4 | 5,9 | 18 | 1893.2 |
| 14 | 1901.7 | 1895,0 | 1895,0 | 1895,0 | 6,7 | 18,6 | 1905.8 |
| 15 | 1913.6 | 1906,7 | 1905,3<br>1908,4 | 1908,4 | 5,2 | 15,2 | 1916.3 |
| 16 | 1923.6 | 1918,6 | 1918,7 | 1918,7 | 4,9 | 15,1 | 1927.0 |
| 17 | 1933.8 | 1927,9<br>1929,9 | 1928,5 | 1928,5 | 5,3 | 15,7 | 1936.5 |
| 18 | 1944.2 | 1940,1 | 1940,0 | 1940,1 | 4,1 | 14,2 | 1947.3 |
| 19 | 1954.3 | 1950,2 | 1949,0 | 1950,2 | 4,1 | 14,7 | 1957.2 |
| 20 | 1964.9 | 1958,0<br>1959,7 | 1959,5 | 1959,5 | 5,2 | 17 | 1968.6 |
| 21 | 1976.5 | 1969,0<br>1971,5 | 1970,6 | 1971,5 | 5,0 | 15,3 | 1979.9 |
| 22 | 1986.8 | 1981,0 | 1981.8 | 1981.8 | 5,0 | 14,6 | 1990.4 |
| 23 | 1996.4 | 1990,8 | 1991,8 | 1991,8 | 4,6 | 15,7 | 1999.7 |
| 24 | 2007,2(?) | 2001,2 | 2001,7 | 2001,7 | 6.0 | | |